\documentclass{aip-cp}

\usepackage[numbers]{natbib}
\usepackage{rotating}
\usepackage{graphicx}
\newcommand{\eqb}{\begin{equation}}
\newcommand{\eqe}{\end{equation}}
\newcommand{\dmb}{\begin{displaymath}}
\newcommand{\dme}{\end{displaymath}}

\newcommand{\eab}{\begin{eqnarray}}
\newcommand{\eae}{\end{eqnarray}}

\begin{document}

\title{The isolated, uniformly moving electron:\\ Selfintersecting SU(2) Yang-Mills center vortex loop and Louis de 
Broglie's hidden thermodynamics}

\author[aff1]{Ralf Hofmann\corref{cor1}}
\eaddress[url]{http://www.thphys.uni-heidelberg.de/~hofmann/}

\affil[aff1]{Institut f\"ur Theoretische Physik, Universit\"at Heidelberg, Philosophenweg 16, 
69120 Heidelberg, Germany}
\corresp[cor1]{Corresponding author: r.hofmann@thphys.uni-heidelberg.de}

\maketitle

\begin{abstract}
We propose that the one-fold selfintersecting center-vortex loop, being the stable excitation in the confining phase 
of SU(2) Yang-Mills thermodynamics of scale $\Lambda\sim 0.5\,$MeV, 
after an electric-magnetically dual interpretation of this theory represents the 
electron/positron. Our argument invokes recent results on the physics of a strongly 
and spherically perturbed 't Hooft-Polyakov monopole, the role of the central spatial region 
in a Harrington-Shepard (HS) (anti)caloron, the latter's deformation towards maximally non-trivial 
holonomy and subsequent dissociation into a pair of 
a screened BPS monopole and antimonopole, the energy-density of the deconfining thermal ground state, and 
the critical temperature $T_c$ for the deconfining-preconfining transition. We estimate the 
typical spatial extent of the selfintersection region and the monopole 
core size, implying that the electron/positron, judged by its Compton wave length, is anything but a point 
particle. Our results support and elucidate 
the ideas of Louis de Broglie on the thermodynamics of an isolated particle.          

\end{abstract}

\section{INTRODUCTION}

In the course of proposing the wave nature of the electron in terms of ``phase harmony" of an internal vibration 
and a wave-like propagation, Louis de 
Broglie stumbled over an apparent contradiction, arising from the distinct Lorentz-boost transformation 
properties of an internal-clock frequency $\omega=\sqrt{1-\beta^2}\omega_0$, transforming in the same way as an 
internal heat $Q=\sqrt{1-\beta^2}Q_0$ or the associated temperature $T=\sqrt{1-\beta^2}T_0$, 
and the total quantum energy $E=\hbar\frac{\omega_0}{\sqrt{1-\beta^2}}\equiv\frac{m_0c^2}{\sqrt{1-\beta^2}}$, 
being the zero component of four-momentum. 
Here $m_0$ denotes the proper mass, $\beta\equiv\frac{v}{c}$ with $v$ being the (linear) velocity 
modulus of the boost and $c$ the speed of light in vacuum, $\hbar$ is Planck's quantum of action, $\omega_0$ represents the 
circular frequency associated with an internal vibration, and $Q_0=m_0c^2$ is the internal heat 
content of the particle in its rest frame. This puzzle was 
addressed by de Broglie in decomposing $E$ as \cite{deBroglieBook1964,deBroglieBook1925} 
\eqb
\label{deBrogliedeco}
E=\hbar\frac{\omega_0}{\sqrt{1-\beta^2}}=\frac{m_0c^2}{\sqrt{1-\beta^2}}=Q+vp\equiv Q+{\cal F}=m_0c^2\sqrt{1-\beta^2}+\frac{m_0 v^2}
{\sqrt{1-\beta^2}}\,,
\eqe 
where the relativistic spatial momentum modulus $p$ is given 
as $p\equiv \frac{m_0 v}{\sqrt{1-\beta^2}}$, and the quantity ${\cal F}=vp$ is interpreted as a 
``pseudo-vis viva" (translational energy) of the particle. In the limit $v\to 0$ the particle $E$ reduces to the rest energy  
$m_0c^2$ which de Broglie considered the temporal mean of a fluctuating
energy being ``the result of the continual energy exchanges between the particle and the hidden thermostat" 
\cite{deBroglieBook1964}. 

The purpose of this brief communication is to propose that the region of 
selfintersection of an SU(2) Yang-Mills center-vortex loop (figure-eight shaped soliton, 
stable excitation in the confining phase, see FIG.\,1)   
\begin{figure}[h]
  \centerline{\includegraphics[width=200pt]{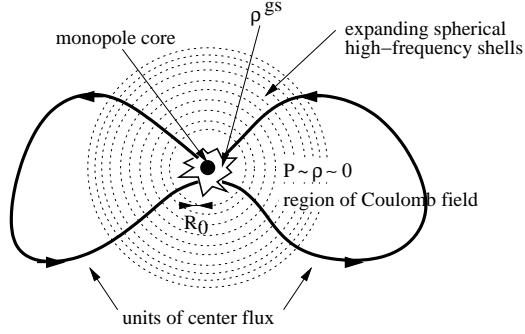}}
  \caption{Schematics of a one-fold selfintersecting center-vortex loop, immersed 
  into the confining phase of SU(2) Yang-Mills theory. SU(2) Yang-Mills theory is to be interpreted in an electric-magnetically dual 
  way, see \cite{HofmannBook2016} and SEC.\,3. Thus, the (vibrating) core of the perturbed magnetic 't Hooft-Polyakov monopole, located within the (fuzzy) selfintersection 
  region of radius $R_0$ representing deconfining thermal-ground-state energy density $\rho^{\rm gs}$, actually represents 
  an electric charge. Also, the magnetic center flux, residing in the two wings of the vortex-loop,  
  actually is an electric one, giving rise to a magnetic moment twice that of a 
  single vortex loop. The concentric circles indicate the dilution 
  of the confining phase by expanding high-frequency shells carried by 
  massive off-Cartan modes of the monopole which also balance the negative pressure of the deconfining phase at the fuzzy phase boundary. 
  These spherical shells are the consequence of 
  (anti)caloron-center induced quantum 
  perturbations of the monopole, and they should enable the penetration of the monopole's 
  electric Coulomb field into the confining phase, compare with SEC.\,4.\label{fig1}}
\end{figure} 
actually satisfies de Broglie's notion of the electron's rest mass $m_0$ being 
the result of quantum induced interactions between a thermal environment, approximated by 
the energy-density and pressure dominating deconfining thermal ground state close to 
the deconfining-preconfining phase boundary at temperature $T_0\sim T_c=\frac{13.87}{2\pi}\Lambda$, $\Lambda$ 
denoting the Yang-Mills scale \cite{HofmannBook2016}, and a topologically stabilised BPS 
monopole. Such a monopole is enabled by the dissociation of an (anti)caloron with 
a large holonomy \cite{Diakonov2004}, in turn, representing a deformed Harrington-Shepard 
(anti)caloron \cite{HS1977} composing the thermal ground state \cite{HofmannBook2016}. For an order-of-magnitude estimate 
of the length scales characterizing this genuine quantum-thermodynamical system the here-considered 
approximations of (i) deconfining SU(2) Yang-Mills 
thermodynamics above but close to $T_0\sim T_c$ being dominated by the thermal ground state, 
(ii) maximally non-trivial (anti)caloron holonomy \cite{Nahm,LeeLu,vanBaalKraanI,vanBaalKraanII}, and (iii) spherically symmetric perturbation of 
the static Bogomolnyi-Prasad-Sommerfield (BPS) monopole \cite{BPS} as in \cite{FodorRacz2004} should suffice.  

This paper is organized as follows. In SEC.\,2 we review some results 
of \cite{FodorRacz2004} on the physics of a strongly excited 't Hooft-Polyakov monopole which are 
relevant for the emergence of the intersection region and the 
leak-out of a Coulomb-like electric field into exterior space ($r>R_0$). 
SEC.\,3 discusses some aspects of SU(2) Yang-Mills thermodynamics 
which, together with the work of \cite{FodorRacz2004}, imply our results of 
SEC.\,4 on the extent of the intersection region and the monopole core. 
In SEC.\,4 we provide a short summary and outlook.          

\section{SEC.\,2: A STRONGLY AND SPHERICALLY PERTURBED 't HOOFT-POLYAKOV MONOPOLE}

If not explicitely stated otherwise, we work in natural units ($\hbar=k_B=c=1$) 
from now on. The normal-mode spectrum of a 
't Hooft-Polyakov, considering small field fluctuations about the static BPS monopole only, 
was investigated in \cite{ForgacsVolkov}. In \cite{FodorRacz2004} a spherically symmetric, strong perturbation of the 't Hooft-Polyakov 
monopole in SU(2) Yang-Mills-Higgs theory was analysed dynamically by virtue of a hyperboloidal conformal 
transformation of the original field equations for the profiles $H(r,t)=h(r,t)/r+H_\infty$ and $w(r,t)$ of the adjoint Higgs and 
off-Cartan fields, respectively. Their results can be summarised as follows. Considering a spherically symmetric, 
localised initial pulse as a strong perturbation of the static BPS monopole, the typical 
dynamical response did not depend on the parameter values of this pulse within a wide range. 
Namely, there are high-frequency oscillations in $w$ which form expanding shells 
decaying in time as $t^{-1/2}$ (the further away from the monopole 
core the shell the higher the frequencies that build it). On the other hand, a localised 
breathing state appears in association with the energy density of the monopole core 
region whose frequency $\omega_0$ approaches the mass $m_w=eH_\infty$ ($e$ denoting the gauge coupling) 
of the two off-Cartan modes in a power-like way in time (natural units): 
\eqb
\label{freutomass}
\omega_0=eH_\infty-C_w t^{-2/3} \ \ \Rightarrow \ \ \lim_{t\to\infty}\omega_0=eH_\infty\,,
\eqe
where $C_w$ is a positive constant. The amplitude of 
the oscillation in energy density decays like $C_a t^{-5/6}$, $C_a$ again denoting  a 
positive constant. Disregarding for now the question what the physics of the exciting initial condition 
is, it thus appears that an internal clock of (circular) frequency $\omega_0\sim m_w$ is run within the core region 
of the perturbed monopole.          

\section{SEC.\,3: SOME ASPECTS OF THE DECONFINING, PRECONFINING, AND CONFINING PHASES IN SU(2) YANG-MILLS THERMODYNAMICS}

To interpret the location of selfintersection within an SU(2) Yang-Mills center-vortex loop as a quantum 
dynamically stabilized region of deconfining thermal ground state, interacting with a 't Hooft-Polyakov 
monopole and immersed in an environent essentially made up of confining phase, 
we need to rely on a few non-perturbative results on SU(2) Yang-Mills 
thermodynamics \cite{HofmannBook2016}.  

\subsection{Deconfining phase}

The deconfining phase of SU(2) Yang-Mills thermodynamics comprises a thermal ground 
state estimate due to HS (anti)calorons. This ground state exhibits a linear-in-$T$ energy density 
\eqb
\label{gsed}
\rho^{\rm gs}=4\pi\Lambda^3 T
\eqe 
and is composed of the densely packed HS-(anti)caloron centers, responsible 
for quantum excitations, and overlapping HS-(anti)caloron peripheries, 
giving rise to (anti)selfdual dipole densities, the 
associated permittivity and permeability and thus the associated propagation 
speed of a wave-like excitation being $T$-independent \cite{HofmannBook2016}. There are off-Cartan massive vector 
modes, which fluctuate quantum-thermodynamically, and a massless 
``photon" mode which propagates in terms of classical electromagnetic waves at 
low frequencies and/or intensities and fluctuates 
quantum-thermodynamically otherwise. A certain class of effective 
radiative corrections collectively describes an ensemble of screened and 
unresolved magnetic monopole-antimonopole pairs, generated by (anti)caloron 
dissociation upon holonomy shift \cite{Nahm,vanBaalKraanI,vanBaalKraanII,LeeLu,Diakonov2004}. The maximum non-trivial 
(anti)caloron holonomy is determined by $A_4(\tau,|\vec{x}|\to\infty)=\pi T t_3$, agreeing on a tr$t^at^b=\frac12\delta^{ab}$ 
normalization of the SU(2) generators $t^a$ ($a=1,2,3$). For a constituent BPS monopole 
this implies $H_\infty=\pi T$. After screening and considering temperature $T$ to be only mildly above $T_c$, 
the associated monopole mass $m_m=\frac{8\pi^2}{e^2}H_\infty$ 
approximately is given as \cite{LudescherKeller2008}
\eqb
\label{m_m}
m_m\sim\frac{8\pi^2}{e^2}H_\infty\sim H_\infty\sim \pi T_c\,,
\eqe
where the plateau value $e=\sqrt{8}\pi$ of the SU(2) coupling 
constant, expressing one-loop thermodynamical selfconsistency of deconfining SU(2) Yang-Mills thermodynamics \cite{HofmannBook2016}, 
was used. For our purposes this is justified because the singularity of $e$ at 
\eqb
\label{TCLamb}
T_c=\frac{13.87}{2\pi}\Lambda\sim \frac{H_\infty}{\pi}
\eqe 
is logarithmically thin only \cite{HofmannBook2016}. Close to $T_c$ 
deconfining SU(2) Yang-Mills thermodynamics is ground-state dominated. Note that 
the radius $R_c$ of the monopole core is about 
\eqb
\label{monocorerad}
R_c\sim H_\infty^{-1}\,.
\eqe

\subsection{Confining and preconfining phase}

The confining phase of an SU(2) Yang-Mills theory exhibits vanishing ground-state pressure and energy 
density at vanishing temperature. As the temperature variable $T$ approaches the Yang-Mills scale 
$\Lambda$ the spectrum of excitations is characterized by an over-exponentially growing density of 
states ($n$-fold selfintersecting center-vortex loops) such that spatially homogeneous 
thermal equilibrium increasingly is invalided \cite{HofmannBook2016}. At $T\sim \Lambda$ a Hagedorn transition occurs, and the intersection regions of center-vortex loops percolate into a condensate of massless magnetic monopoles/antimonopoles, forming 
the new thermal ground state of the preconfining phase \cite{HofmannBook2016}. This ground state 
dominates the entire thermodynamics due to the Meissner massiveness of the gauge mode which is massless in the 
deconfining phase. The latter sets in at $T_c$ \cite{HofmannBook2016}.       

\subsection{Electric-magnetically dual interpretation 
of charges in SU(2) Yang-Mills theory\label{EMD}}

In units were $c=\epsilon_0=\mu_0=1$ ($\epsilon_0$ and $\mu_0$ denoting the electric permittivity and magnetic permeability of 
the classical electromagnetic vacuum in SI units) the QED fine structure 
constant $\alpha$, which is unitless in any system of units, 
reads
\eqb
\label{alpha}
\alpha=\frac{Q^2}{4\pi\hbar}\,.
\eqe
On the other hand, the Yang-Mills coupling $e$ was argued to 
be $e=\frac{\sqrt{8}\pi}{\sqrt{\hbar}}$ almost everywhere 
in the deconfining phase \cite{KavianiHofmann2012}, based on the observation in \cite{HerbstHofmann2004} that 
only a small range of (anti)caloron radii, centered about 
the spatial coarse-graining cutoff $|\phi|^{-1}$, contributes to the 
emergent thermal ground state. This value of $e$ implies the (anti)caloron action to 
be $\hbar$. Now, for $\alpha$ to be unitless it follows that $Q\propto 1/e$, meaning that an electric 
charge in the real world is represented by a magnetic charge w.r.t. to the Cartan subalgebra in 
the SU(2) Yang-Mills theory. In particular, a magnetic monopole in the deconfining phase of an 
SU(2) Yang-Mills theory thus is to be interpreted as an electric charge in our world.

\section{SEC.\,4: EXTENDED ELECTRIC CHARGE AND MAGNETIC MOMENT}

Let us now exploit the facts introduced in the previous sections to make contact with de Broglie's 
hidden thermodynamics of the isolated electron and to infer typical length scales governing the selfintersection 
region of the associated center-vortex loop in SU(2) Yang-Mills theory. 

\noindent First, according to Eq.\,(\ref{gsed}) the energy density of 
the deconfining ground state, which dominates SU(2) Yang-Mills thermodynamics for $T$ close to $T_c$,  
is {\sl linear} in $T$ ($T=\sqrt{1-\beta^2}T_0$), supporting the interpretation that the internal heat $Q$ is due to 
the ``thermostat" interacting with the 
particle: $Q=\sqrt{1-\beta^2}Q_0$. 

\noindent Second, according to Eq.\,(\ref{freutomass}) one may express the rest energy/mass $m_0$, associated with the intersection region, as
\eqb
\label{massele}
m_0\sim e H_\infty\sim \sqrt{8}\pi H_\infty\,.
\eqe

\noindent On the other hand, due to energy conservation exhibited by the quantum interaction of the deconfining 
thermal ground state with the isolated monopole -- induced by a sequence of perturbations issued indeterministically by (anti)caloron centers \cite{Entropy2016} --, we may write (compare with Eqs.\,(\ref{gsed}),(\ref{m_m}),(\ref{TCLamb}))
\eqb
\label{masseleqi}
m_0=m_m+E_0\sim H_\infty+\frac43\pi R_0^3\,\rho^{\rm gs}\sim H_\infty\left(1+
\frac{128\pi}{3}\left(\frac{R_0}{13.87}\right)^3 H^3_\infty\right)\,,
\eqe
where $R_0$ denotes the spatial radius of a ball-like (yet fuzzy) region of 
deconfining phase, see Fig.\,\ref{fig1}. Equating the right-hand sides of Eqs.\,(\ref{massele}), (\ref{masseleqi}) 
and solving for $R_0$, 
we arrive at
\eqb
\label{R0}
R_0\sim 13.87\left(\frac{3}{128\pi}\left(\sqrt{8}\pi-1\right)\right)^{1/3}\,H^{-1}_\infty\sim 5.4\,H^{-1}_\infty\,.
\eqe
Comparing this to the Compton wave length $\lambda_C=m_0^{-1}$, we have
\eqb
\label{CWL}
R_0\sim 46.9\,\lambda_C\sim 1.1\times 10^{-10}\,\mbox{m}\,.
\eqe
Thus, $R_0$ is about twice as large as the Bohr radius $a_0$  which, in turn, is about $2.7(=5.4/2)$ 
times larger than the radius of the monopole core $R_c$. Therefore, it would be utterly incorrect to consider 
the electron, if realized in the manner as proposed above, to be a point particle. 
The reason why scattering experiments do not reveal an inner structure is that 
{\sl quantum thermodynamics}, being the state of maximum entropy taking place within radius $R_0$, does not 
represent any discernible structure. As discussed in SEC.\,2, based on the important work \cite{FodorRacz2004}, 
high-frequency 
oscillations in the profile function $w$ of the off-Cartan gauge fields, belonging to the perturbed monopole, develop 
expanding shells. If these shells are considered to penetrate the 
confining phase ($r>R_0$), such that the static Coulomb field of the monopole, arising from a temporal average 
over many cycles of the monopole core-vibration, can actually permeate this 
phase, then we may estimate the Coulomb-field correction $\Delta m_0$ to the ``thermodynamical" mass $m_0$ as
\eqb
\label{Coulomb} 
\Delta m_0=\int_{R_0}^{\infty} dr\,\frac{r^2}{r^4}=R_0^{-1}\sim\frac{1}{5.4} H_\infty\ll 
\sqrt{8}\pi H_\infty=m_0\,\ \ \ \mbox{or} \ \ \ m_0\sim 48\,\Delta m_0\,.
\eqe
This is a $\sim$2\% correction only, much in contrast to the classical 
electron radius defining the entire rest mass $m_0$ in terms of Coulomb energy.  

\noindent Finally, we would like to mention that, in assigning half a Bohr 
magneton to the magnetic moment carried by a single center-vortex loop, a $g$-factor of 
two naturally arises by the electron being composed of two such center-vortex 
loops by virtue of the region of selfintersection, see Fig.\,\ref{fig1}. 
 
\section{SUMMARY}

In this contribution we have, based on a recent numerical investigation of an isotropically perturbed 
't Hooft-Polyakov monopole \cite{FodorRacz2004}, the constituents / structure of the 
deconfining thermal ground state, and the nature of 
excitations in the confining phase of SU(2) Yang-Mills thermodynamics \cite{HofmannBook2016}, 
elucidated the assertion made by L. de Broglie \cite{deBroglieBook1925,deBroglieBook1964} that, based on distinct Lorentz boost transformation 
properties of frequency $\omega_0$ and energy $E$, both associated with 
the rest mass $m_0$, the electron's charge and mass represent a spatial region of 
active quantum thermodynamics in SU(2) Yang-Mills theory. Our main result is that the radial extent $R_0$ of this region is comparable 
to typical atomic radii with the radial extent $R_c$ of the electric charge being about a factor of three smaller: $R_0$: $R_c\sim \frac13 a_0$, $a_0$ being the Bohr radius. The discussion of the present paper is restricted to a 
uniformly moving electron / positron. How such a system responds 
to acceleration by external forces is, as of yet, unclear because this would necessitate to solve the 
field equations under essentially relaxed symmetry assumptions.       

\section{ACKNOWLEDGMENTS}
We thank Manfried Faber and Steffen Hahn for useful conversations. 


\nocite{*}
\bibliographystyle{aipnum-cp}%

\bibliography{RH}

\end{document}